\documentclass[journal]{IEEEtran}

\usepackage{cite}
\usepackage{amsmath,amssymb,amsfonts}
\usepackage{graphicx}
\usepackage{booktabs}
\usepackage{array}
\usepackage{siunitx}
\usepackage{dblfloatfix}
\usepackage{hyperref}
\hypersetup{hidelinks}

\setkeys{Gin}{width=\columnwidth,keepaspectratio}
\graphicspath{{figs/}}

\title{Quantum-Enhanced Analysis and Grading of Vocal Performance}

\author{Rohan Agarwal%
\thanks{R. Agarwal is with Monta Vista High School and De Anza College (dual enrollment), Cupertino, CA, USA
(\href{mailto:agarwalrohan2@student.deanza.edu}{agarwalrohan2@student.deanza.edu}).}}

\begin{document}
\maketitle

\begin{abstract}
Vocal singing is a profoundly emotional art form possibly predating spoken language, yet evaluating a vocal track remains a subjective and specialized task. Meanwhile, quantum computing shows promise to bring about significant advances in science and art. This study introduces \textit{QuantumMelody}, a quantum-enhanced algorithm to evaluate vocal performances through objective metrics. QuantumMelody begins by collecting a comprehensive array of classical acoustic and musical features including pitch contours, formant frequencies, Mel-spectrograms, and dynamic ranges. These features are divided into three musically categorized groups, converted into scaled angles based on statistical metrics, and then encoded into specific quantum rotation gates. Each qubit group is entangled internally, followed by intergroup entanglement, thus exploring subtle, non-linear relationships within and across feature sets. The resulting quantum probability distributions and classical features are used to train a neural network, combined with a spectrogram transformer to holistically grade each recording on a 2--5 scale. Key difference metrics like the Jensen--Shannon distance and Euclidean measures of scaled angles are used to enable nuanced comparisons of different recordings. Furthermore, the algorithm uses classical music-based heuristics to provide targeted suggestions to the user for various aspects of vocal technique. On a dataset of 168 labeled 20\,s vocal excerpts, QuantumMelody achieves \SI{74.29}{\percent} agreement with expert graders. The circuits are simulated; we do not claim hardware speedups, and results reflect a modest, single-domain dataset. We position this as an applied audio-signal-processing contribution and a feasibility step toward objective, interpretable feedback in singing assessment. 
\end{abstract}


\section{Introduction}
Assessing the quality of a singing performance is a very subjective and specialized task. This is especially important in music schools, where hundreds of student recordings might need review each week and become a severe bottleneck for music teachers. The motivation for this research is rooted in the need to offer a reliable automated system for feedback with clear, consistent pointers for improvement.

Prior research in singing voice analysis provides a foundation for an objective approach. For example, humans tend to judge performances by concrete aspects such as pitch accuracy, stability of tone, timing, and vocal timbre, even if their overall impressions are subjective. Work by Ghisingh \textit{et al.} (2017) \cite{ghisingh2017tencon} analyzed Indian classical singing by isolating the vocal track from background music and examining acoustic production features like pitch and root-mean-square energy. Other researchers have built systems to grade singing on metrics including intonation correctness, rhythm consistency, and vibrato usage. For instance, Liu and Wallmark (2024) \cite{liu2024peking} trained machine learning models on annotated singer characteristics to classify timbre and technique attributes in traditional opera singing. Meanwhile, recent surveys by Hashem \textit{et al.} (2023) \cite{hashem2023ser} document speech emotion recognition systems that infer emotions from voice signals via deep learning. These advances illustrate that modern AI and machine learning (ML) can decipher subtle information from vocal audio, whether it be emotional tone or singing skill.

While classical ML approaches for voice grading are promising, nascent technologies like quantum computing open new frontiers for audio analysis. Quantum computing's ability to represent and entangle high-dimensional data offers a novel way to capture the complex interdependencies of musical features. Kashani \textit{et al.} (2022) \cite{kashani2022qft} implemented a note detection algorithm based on the Quantum Fourier Transform (QFT). Miranda \textit{et al.} (2021) \cite{miranda2021qnlp} envisioned quantum computing's impact on music technology and introduced frameworks for quantum music intelligence. G\"{u}nd\"{u}z (2023) \cite{gunduz2023entropy} used information-theoretic metrics like Shannon entropy to quantify musical complexity. These developments set the stage for a hybrid approach to the long-standing challenge of vocal performance evaluation.

\textit{QuantumMelody} is proposed as a solution that combines robust audio feature analysis with quantum computing to achieve consistent and insightful vocal grading. First, the algorithm extracts a feature vector from raw singing audio, encompassing both traditional and innovative descriptors of vocal quality. These features capture pitch and intonation accuracy, frequency stability (jitter) and amplitude stability (shimmer), LUFS energy (loudness and dynamics), and timbral characteristics such as MFCCs and formant frequencies. Expressive elements like vibrato extent and rate are also included.
\section{Materials and Methods}
\subsection{Dataset and Preprocessing}
We collect \SI{20}{\second} vocal recordings in seven Hindustani ragas, with 42 labeled samples each for ratings of 2, 3, 4 and 5 across the seven ragas, for a total of 168 samples. All audio is converted to mono and resampled to \SI{22050}{Hz} using Librosa's high-quality resampler. We then apply denoising and drone removal by attenuating low-frequency harmonics corresponding to a tanpura drone via a narrow band-stop filter around the drone frequency. This is followed by harmonic--percussive separation to extract the pure vocal track. After filtering, we use a short-time Fourier transform (STFT) \cite{ghisingh2017tencon} to verify that the drone, percussion, and any low-frequency noise are removed while preserving the singer's voice.

\subsection{Feature Extraction}
We extract a suite of time-domain and frequency-domain features for each recording as follows.
\paragraph{Pitch deviation (cents)} Let the instantaneous fundamental frequency be $F_0[n]$ and the tonic be $F_{\mathrm{ref}}$. The per-frame pitch deviation in cents is
\begin{equation}
\Delta[n] = 1200 \log_2\!\left(\frac{F_0[n]}{F_{\mathrm{ref}}}\right).
\label{eq:pitch_dev}
\end{equation}
We compute the average absolute pitch deviation $\tfrac{1}{N}\sum_{n=1}^N |\Delta[n]|$, which quantifies pitch stability.

\paragraph{Jitter} With successive pitch periods $T_i = 1/F_0[i]$, the local jitter is
\begin{equation}
J_{\mathrm{local}} = \frac{1}{M-1}\sum_{i=1}^{M-1}\frac{|T_{i+1}-T_i|}{T_i},
\end{equation}
reported as a percentage.

\paragraph{Shimmer} With glottal pulse peak amplitudes $A_i$, the local shimmer is
\begin{equation}
S_{\mathrm{local}} = \frac{1}{M-1}\sum_{i=1}^{M-1}\frac{|A_{i+1}-A_i|}{A_i}.
\end{equation}

\paragraph{Loudness (LUFS) and RMS energy} ITU-R BS.1770 loudness is approximated by
\begin{equation}
L_{\mathrm{LUFS}} = -0.691 + 10\log_{10}\!\left(\frac{1}{N}\sum_{n=1}^{N} w[n]^2\right),
\end{equation}
where $w[n]$ is the K-weighted waveform. We also compute
\begin{equation}
E_{\mathrm{RMS}} = \sqrt{\frac{1}{N}\sum_{n=1}^{N} x[n]^2}.
\end{equation}

\paragraph{Tone-to-Noise Ratio (TNR)} Using Parselmouth, we estimate
\begin{equation}
\mathrm{TNR_{dB}} = 10\log_{10}\!\left(\frac{P_{\mathrm{tone}}}{P_{\mathrm{noise}}}\right).
\end{equation}

\paragraph{MFCCs} From log-Mel energies $S_m$, the $k$th MFCC is
\begin{equation}
\mathrm{MFCC}_k = \sum_{m=1}^{M} \log(S_m)\cos\!\left[\frac{\pi k}{M}\Big(m-\tfrac{1}{2}\Big)\right].
\end{equation}

\paragraph{Zero-Crossing Rate (ZCR)} The short-term zero-crossing rate per frame is
\begin{equation}
Z = \frac{1}{N-1}\sum_{n=1}^{N-1}\mathbf{1}\{x[n]x[n+1] < 0\}.
\end{equation}

\paragraph{Spectral centroid}
\begin{equation}
C = \frac{\sum_f f\,|X(f)|}{\sum_f |X(f)|}.
\end{equation}

\paragraph{Spectral bandwidth}
\begin{equation}
B = \sqrt{\frac{\sum_f (f-C)^2 |X(f)|}{\sum_f |X(f)|}}.
\end{equation}

\paragraph{Spectral flatness}
\begin{equation}
F = \frac{\exp\!\big(\frac{1}{K}\sum_{k=1}^{K}\ln P_k\big)}{\frac{1}{K}\sum_{k=1}^{K} P_k},
\end{equation}
where $P_k = |X(f_k)|^2$.

\paragraph{Formants F1--F3} Using Burg LPC we estimate the first three formants per voiced frame and take their means (Hz).

\paragraph{Vibrato extent and rate} For the pitch contour $F_0[n]$, the vibrato extent (in cents) is
\begin{equation}
E_v = 1200 \log_2\!\left(\frac{\max_n F_0[n]}{\min_n F_0[n]}\right),
\end{equation}
and the rate is the dominant modulation frequency of $F_0[n]$ (Hz).
\vspace{0.25em}

\subsection{Angle Scaling}
After computing raw features, we scale them to angles in $[0,2\pi]$ to map to the Bloch sphere and construct the quantum circuit. Representative mappings include:
\begin{align}
\theta_{\text{pitch}} &= 2\pi\cdot \frac{1}{1 + e^{-\frac{(d-d_0)}{k}}},\\
\theta_{\text{jitter}} &= 2\pi\cdot \frac{1}{1 + e^{-a(J-J_0)}},\\
\theta_{\text{tempo}} &= 2\pi \cdot \tanh\!\left(\frac{T_{\text{rel}}}{r}\right),\\
\theta_{\text{shimmer}} &= 2\pi\cdot \frac{1}{1 + e^{-b(S-S_0)}},\\
\theta_{\text{LUFS}} &= 2\pi\cdot \frac{L - L_{\min}}{L_{\max} - L_{\min}},\\
\theta_{\sigma\text{LUFS}} &= 2\pi\cdot \frac{\sigma_{\text{LUFS}}}{\sigma_{\max}},\\
\theta_{\text{MFCC}} &= 2\pi\cdot \frac{\ln(1+M)}{\ln(1+M_{\max})},\\
\theta_{\text{ZCR}} &= 2\pi\cdot \frac{\min(ZCR,0.2)}{0.2},\\
\theta_{\text{TNR}} &= 2\pi\cdot \frac{T_{\max} - \mathrm{TNR}}{T_{\max} - T_{\min}}.
\end{align}
Here, $T_{\max}$ and $T_{\min}$ are the empirical bounds for TNR (dB); analogous symbols are used for other features. Unless noted, scaling parameters ($d_0,k,a,b,S_0,r,\sigma_{\max},M_{\max}$) are fixed from the training set using robust bounds (5th–95th percentile or empirical min/max as indicated) and remain constant during evaluation.

\begin{table}[!t]
\caption{Features, rotation group, and empirical scaling bounds.}
\label{tab:bounds}
\centering
\begin{tabular}{@{}p{0.34\linewidth}p{0.27\linewidth}p{0.14\linewidth}p{0.14\linewidth}@{}}
\toprule
Feature & Rotation Group & Min & Max\\
\midrule
Average pitch deviation (cents) & $R_x$ (pitch stability) & 0 & 1431.7\\
Average jitter & $R_x$ (pitch stability) & 0 & 0.3278\\
Std.\ dev.\ tempo (BPM) & $R_x$ (rhythm) & 30 & 180\\
Average shimmer & $R_y$ (dynamics) & 0 & 1.1735\\
Mean LUFS energy (dB) & $R_y$ (dynamics) & $-60$ & $-10$\\
Std.\ dev.\ LUFS energy & $R_y$ (expression) & 1 & 12\\
Std.\ dev.\ MFCC (timbre) & $R_z$ (timbre) & 0 & 0.25\\
Zero-crossing rate & $R_z$ (clarity) & 0.01 & 0.12\\
Mean tone-to-noise ratio (TNR, dB) & $R_z$ (clarity) & 5 & 30\\
\bottomrule
\end{tabular}
\end{table}

\subsection{Quantum Circuit Architecture}
We construct a 9-qubit circuit in Qiskit. All qubits are initialized with a Hadamard layer $H^{\otimes 9}$, then receive rotations based on grouped features: qubits 0--2 receive $R_x(\theta_i)$ (pitch-related angles), 3--5 receive $R_y(\theta_i)$ (dynamics), and 6--8 receive $R_z(\theta_i)$ (timbre). We apply intra-group CNOTs (0$\rightarrow$1, 1$\rightarrow$2; 3$\rightarrow$4, 4$\rightarrow$5; 6$\rightarrow$7, 7$\rightarrow$8) and cross-group CNOTs (2$\rightarrow$3, 1$\rightarrow$4, 0$\rightarrow$6, 5$\rightarrow$7), then measure all qubits with \num{8192} shots on the \textit{Qiskit simulator} to obtain measurement probabilities.

\subsection{Hybrid Neural Network}
We combine an Audio Spectrogram Transformer (AST) \cite{gong2021ast} fine-tuned on Mel-spectrograms with parallel MLPs over classical features and quantum-derived features. The concatenated embeddings are fed to a fully-connected head that predicts a categorical grade (2--5).

\subsection{Comparison Metrics and Environment}
For two recordings with quantum probability distributions $P$ and $Q$, the Jensen--Shannon divergence is
\begin{equation}
D_{\mathrm{JS}}(P\Vert Q) = \tfrac{1}{2}D_{\mathrm{KL}}(P\Vert M) + \tfrac{1}{2}D_{\mathrm{KL}}(Q\Vert M),\quad M=\tfrac{1}{2}(P+Q).
\end{equation}
where $D_{KL}(\cdot\Vert\cdot)$ denotes the Kullback--Leibler divergence.
We also compute the Euclidean distance between scaled-angle vectors $\boldsymbol{\theta}^{(1)}$ and $\boldsymbol{\theta}^{(2)}$:
\begin{equation}
d=\sqrt{\sum_{i=1}^{n}\bigl(\theta_i^{(1)}-\theta_i^{(2)}\bigr)^2}.
\end{equation}
All code is written in Python~3.12 using Librosa, Parselmouth, Qiskit, PennyLane and PyTorch on macOS.

\subsection{Ethics and Reproducibility}
All participants provided informed consent for recording and research use of their vocal audio. Data are anonymized; no identifying metadata are released.

Audio is mono at \SI{22.05}{\kilo\hertz}. Feature set: \(F_0\) deviation (cents), jitter, shimmer, LUFS/RMS, TNR, MFCCs (first 13, \(\mu/\sigma\)), ZCR, spectral centroid/bandwidth/flatness, formants, and vibrato extent/rate. The quantum circuit configuration and training code are available on request; a companion repository with scripts and hyperparameters will be posted after this preprint is announced.
\section{Results and Discussion}
\begin{figure}[!t]
\centering
\includegraphics{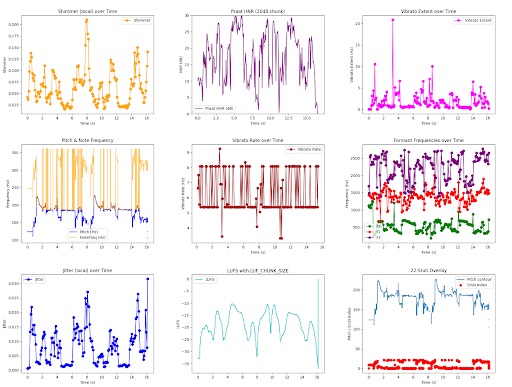}
\caption{Classical metrics used as inputs to the combined classical--quantum model.}
\label{fig:1}
\end{figure}

The hybrid classical–quantum framework processed each vocal recording in approximately one minute on average (\textit{Qiskit simulator}). We do not claim hardware speedups; real-time performance on quantum hardware is outside our scope. Fig.~\ref{fig:1} shows the classical metrics captured for each recording and used as inputs to the combined model.

\begin{figure}[!t]
\centering
\includegraphics{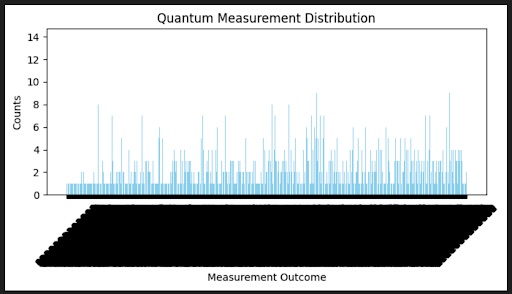}
\caption{Quantum measurement distribution over bitstrings.}
\label{fig:2}
\end{figure}

Quantum measurement distributions were calculated for each circuit using \num{8192} shots (Fig.~\ref{fig:2}). Most recordings did not display a single spike but showed a clear shape, indicating intertwined features.

\subsection{Improvement over Classical Methods}
\begin{figure}[!t]
\centering
\includegraphics{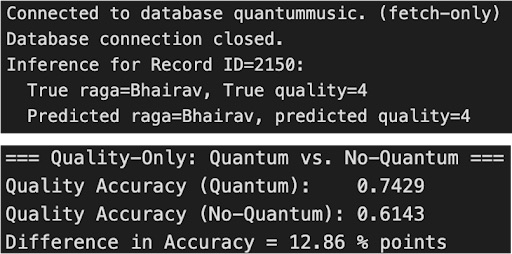}
\caption{Quality-only comparison: quantum-enhanced vs.\ classical baseline.}
\label{fig:3}
\end{figure}

The baseline included classical features only (pitch dev., jitter/shimmer, MFCC stats, TNR, LUFS, ZCR, formants) with an MLP; the hybrid model adds AST embeddings + quantum measurement probabilities. We used an 80/20 stratified split with raga/label balance and no speaker leakage. We report \emph{agreement}, defined as the percentage of samples whose predicted label exactly matches the expert-provided grade. We cast 2--5 grading as a four-class classification task and report agreement with expert graders. Compared to the classical-only baseline, the hybrid model achieved a +\num{12.86}-point absolute improvement in \emph{agreement with expert graders}, reaching \SI{74.29}{\percent} versus \SI{61.43}{\percent} for the classical system (Fig.~\ref{fig:3}). Given n=168, estimates have non-trivial variance; we report point estimates here.

\subsection{Student vs.\ Master Comparison}
\begin{figure}[!t]
\centering
\includegraphics{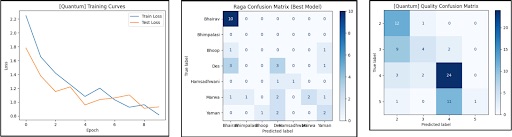}
\caption{Hybrid AST model training curves and confusion matrices.}
\label{fig:4}
\end{figure}

\begin{figure}[!t]
\centering
\includegraphics{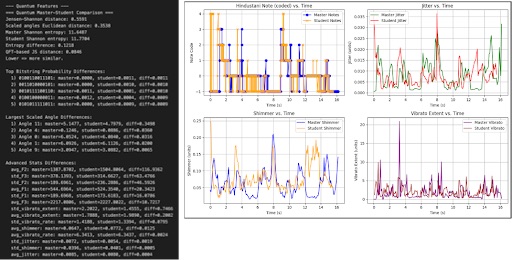}
\caption{Student vs.\ master comparison across selected features.}
\label{fig:5}
\end{figure}

Divergence measures—Jensen–Shannon and Euclidean—computed on quantum-encoded feature vectors \emph{qualitatively aligned} with perceptual discrepancies between student and teacher recordings. Analysis showed pitch deviations of 15--25 cents from master tracks (ideal $<$10 cents), LUFS differences up to \SI{3}{dB}, and TNR values of \SIrange{12}{18}{dB} for students vs.\ $> \SI{20}{dB}$ for teachers (Fig.~\ref{fig:5}). Training curves and confusion matrices for the AST model are in Fig.~\ref{fig:4}.

\section{Conclusion}
We presented \textit{QuantumMelody}, a hybrid method that encodes grouped vocal features in a compact quantum circuit and fuses the circuit’s measurement probabilities with spectrogram-transformer embeddings to estimate a 2--5 grade and surface technique-level feedback. The circuit uses nine qubits with $R_x$, $R_y$, and $R_z$ encodings aligned respectively with pitch stability, dynamics, and timbre, with intra- and inter-group entanglement to model cross-domain interactions.

On \num{168} labeled \SI{20}{\second} excerpts, the hybrid model attains \SI{74.29}{\percent} agreement with expert graders, a +\num{12.86}-point improvement over a classical-features baseline. All quantum results are produced \emph{on a laptop-class Qiskit simulator}; we do not claim hardware speedups, and behavior on current NISQ devices may differ. Given the modest, single-domain dataset, these findings should be interpreted as a feasibility demonstration within applied audio signal processing.

Overall, the approach provides objective, interpretable indicators for vocal technique without relying on quantum hardware performance claims. This work is most appropriately read within applied audio signal processing.

\bibliographystyle{IEEEtran}
\bibliography{references}

\end{document}